# Quantized one-dimensional edge channels with strong spin-orbit coupling in 3D topological insulators

Christoph Kastl,[1,2] Paul Seifert,[1,2] Xiaoyue He,[3] Kehui Wu,[3] Yongqing Li,[3] Alexander Holleitner[1,2*]

[1] *Walter Schottky Institut and Physics department, Technical University of Munich, Am Coulombwall 4a, 85748 Garching, Germany.*

[2] *Nanosystems Initiative Munich (NIM), Schellingstr. 4, 80799 München, Germany.*

[3] *Institute of Physics, Chinese Academy of Sciences, Beijing 100190, China.*

*holleitner@wsi.tum.de

**A strong coupling between the electron spin and its motion is one of the prerequisites of spin-based data storage and electronics.[1,2] A major obstacle is to find spin-orbit coupled materials where the electron spin can be probed and manipulated on macroscopic length scales, for instance across the gate channel of a spin-transistor.[3,4] Here, we report on millimeter-scale edge channels with a conductance quantized at a single quantum $1 \cdot e^2/h$ at zero magnetic field. The quantum transport is found at the lateral edges of three-dimensional topological insulators made of bismuth chalcogenides.[5,6,7] The data are explained by a lateral, one-dimensional quantum confinement of non-topological surface states with a strong Rashba spin-orbit coupling.[8] This edge transport can be switched on and off by an electrostatic field-effect. Our results are fundamentally different from an edge transport in quantum spin Hall insulators and quantum anomalous Hall insulators.[9,10]**

Bismuth chalcogenides such as $Bi_2Te_3$, $Bi_2Se_3$ and $BiSbTe_3$ were formerly studied as narrow band gap semiconductors for thermoelectric applications.[11] Nowadays, these compounds are better known as three-dimensional topological insulators, which exhibit two-dimensional gapless surface states with spin-helical Dirac dispersion.[5,6,12] The peculiar helical structure of the surface states is a direct consequence of a strong spin-orbit interaction inherent to heavy metal compounds. Additionally, bismuth chalcogenides have non-topological surface states with a large Rashba spin-orbit interaction due to a structure inversion asymmetry.[8,13,14] In principle, the spin-orbit interaction offers the unique possibility to control the electron spin by means of electric fields.[1] Unfortunately, the coupling between electron motion and spin tends to randomize spin information very quickly by momentum scattering.[15] In this respect, quasi one-dimensional systems are particularly interesting, since the momentum space for scattering is reduced.[16,17] An external magnetic field can further open a gap in the excitation spectrum and a helical regime can be established, where counter-propagating electrons have opposite spin.[4] Such helical systems are expected to enable novel types of information processing, such as topological quantum computing via Majorana fermions.[18,19,20] By a combined scanning photocurrent/voltage microscopy,[21,22,23] we show the existence of one-dimensional channels at etched lateral boundaries of bismuth chalcogenide films. By a field effect gating, the edge channels are switched on and off. We determine a conductance quantization at $1 \cdot e^2/h$ at zero magnetic field implicating that the time-reversal symmetry is preserved; a regime which cannot be reached by quantum anomalous Hall insulators.[10] Furthermore, we observe the quantized conductance on a millimetre scale. This distinguishes our results from the quantum spin Hall insulator where spin dephasing limits its observation to microscale length scales.[9,24] Our numerical simulations suggest that the measured quantum transport is induced by a dominating strong spin-orbit coupling in the edge states.[25,26] Hence, bismuth chalcogenides offer a promising platform to realize one-dimensional electron systems with a strong spin-orbit coupling, which are scalable, electrically tunable and which do not require magnetic fields.

Films of $Bi_2Se_3$ ($BiSbTe_3$) with a thickness of 10 nm (15 nm) are MBE grown on insulating (111) $SrTiO_3$ and patterned into Hall bar circuits (Fig. 1a and Methods).[21,22,23] We scan a laser beam across the circuit and measure the photocurrent between source and drain $I_{photo}(x,y)$ at each excitation position at zero bias (Fig. 1a). Simultaneously, a differential amplifier senses the photovoltage $V_{photo}(x,y) = V_{voltage} - V_{drain}$ between the drain ($V_{drain}$) and a further voltage probe ($V_{voltage}$). The charge density is adjusted via a global backgate voltage $V_{gate}$.[22] All contacts except source and drain are floating. Figure 1b shows a photocurrent map of the area which is highlighted as a dotted rectangle in Fig. 1a. At a negative $V_{gate}$, the signal is dominated by sub-micron photocurrent patterns, which can be understood as a thermoelectric photocurrent induced by potential fluctuations at the bottom surface of the $Bi_2Se_3$-films (Fig. 1b).[22] At a positive $V_{gate}$, an additional, ~40 times larger photocurrent shows up at the lateral circuit edges (Fig. 1c). The concurrently recorded photovoltage map is depicted in Fig. 1d. Strikingly, the photoresponse at the edges persists over hundreds of micrometers (Supplementary Figure 1), which is orders of magnitude larger than the diffusion length of optically excited charge carriers in the materials.[27]

Generally, the quantum conductance of an ideal 1D, non-interacting electron system is an integer multiple of $e^2/h$

$$G = n \cdot g \cdot e^2/h, \quad (1)$$

with $n$ the number of modes. At zero magnetic field, the spin degeneracy $g$ equals 2. We define the photoconductance $G_{photo} = I_{photo}/V_{photo}$ and calculate the ratio $G_{photo}$ along the edge between source and drain marked by triangles in Figs. 1c and 1d. We find that $G_{\text{photo}}$ is constant along the edges (Fig. 1e) and quantized at a mean value of $G_{\text{photo}} = (1.03 \pm 0.005) \cdot e^2/h$ with a full width at half maximum (FWHM) $\Delta G_{\text{photo}} = 0.091\ e^2/h$ (Fig. 1f). We determine the noise of the current and voltage signals, and calculate the expected FWHM of the histogram to be $\Delta G_{\text{photo}}^{\text{calc}} = 0.096\ e^2/h$ (blue line in Fig. 1f, and Supplementary Note 1). In other words, the measured conductance quantization is only limited by the apparent signal-to-noise ratio. For BiSbTe$_3$, we observe equivalent edge channels with a conductance quantization of $G = 1.03\ e^2/h$ with a full width at half maximum (FWHM) $\Delta G_{\text{photo}} = 0.12\ e^2/h$ (Supplementary Figure 2).

Next, we elucidate how the 1D channels form at the lateral edges of the circuit. The solid black line in Fig. 2a depicts the source-drain conductance $G_{\text{SD}}(V_{\text{gate}})$ without laser illumination ($P_{\text{laser}} = 0$ nW). Under a homogenous illumination of the sample, $G_{\text{SD}}$ surprisingly reduces for $V_{\text{gate}} \geq 20$ V and it develops into a plateau for an increased laser intensity. Our photocurrent/voltage spectroscopy reveals that the plateau is a signature of macroscopic edge currents. For $V_{\text{gate}} \geq 20$ V, the dominating optoelectronic response is localized at the lateral edges of the circuits (Fig. 1c). In contrast, for $V_{\text{gate}} = -30$ V (Fig. 1b), the one-dimensional channels are absent. From the fact, that a Fermi level tuning switches the one-dimensional edge transport on and off, we can exclude two-dimensional topological surface states as the transport-carrying states. They extend continuously across the whole bulk band gap. Accordingly, they cannot simply be switched on by a positive gate voltage, as we find experimentally (Fig. 1b vs. 1c). Instead, we propose that the edge channels result from a lateral confinement of topologically trivial states with a strong spin-orbit coupling.

The surface of a Bi$_2$Se$_3$ film is only chemically stable under ultra-high vacuum conditions. Inevitably, upon exposure to atmospheric conditions, e.g. during microlithogra-phy processes, a two-dimensional electron inversion layer (2D-IL) forms.[14,28] This topologically trivial surface state coexists with the topological state.[8] This arrangement can be understood by a surface doping and a van der Waals gap expansion of the outermost quintuple layers.[13,14,28] Furthermore, plasma etching induces defects at unprotected surfaces.[29] For the $Bi_2Se_3$ and $BiSbTe_3$-circuits, such defects occur at the vertical facets of the lateral edges resulting in a lateral band bending on the order of tens of meV,[12] as recently demonstrated by us.[22] The apparent width of the transport channels in Figs. 1c and 1d is given by the diffraction limited spot size (~µm). However, the lateral confinement length can be estimated to be as short as several nanometers due to the screening of the coexisting topological surface states.[30] Fig. 2b sketches a corresponding band diagram at the edges of the Hall bars including the two-dimensional inversion layer (2D-IL)[14,28] and the defect states occurring at the vertical facets.[12,22] An optical excitation creates electron and hole pairs across the band gap. At the edges, the electrons can relax via intermediate states into the localized defect states, whereas the holes tend to drift away from the edges. With an assumed optically induced Fermi level pinning at the edges, laterally extended one-dimensional quantum wire states form along the overall boundary of the circuit (Figs. 1c and 1d). The photoexcitation locally increases the chemical potential at the laser position, either by a direct injection of photogenerated charge carriers into the edge channels or by a capacitive coupling to the edge channels via the population of trap states.[31] To corroborate our trap state model, we show the laser intensity dependence of $I_{photo}$ in the edge current dominated regime ($V_{gate} = +80$ V, Fig. 2c) and in the potential fluctuation dominated regime ($V_{gate} = -40$ V, Fig. 2d). In the latter, the photocurrent scales linearly with laser intensity,[21] whereas in the former, the photocurrent shows a sublinear dependence on the optical power (Fig. 2c). A power law fit $I_{photo} = I_0 \cdot P_{laser}^p$ to the data yields $p = 0.7$. For $BiSbTe_3$, we find $p =$ 0.6 (Supplementary Figure 3). An exponent $0.5 \leq p < 1$ is a clear characteristic for a photocurrent generation involving trap states.[32]

We now address the question, why the conductance is quantized at $1 \cdot e^2/h$ rather than at $2 \cdot e^2/h$. Figure 3a depicts numerical simulations of quasi one-dimensional edge modes using the Kwant code[33] (Supplementary Note 2 and Supplementary Figure 4). In particular, the figure shows the parabolas of the first and second one-dimensional subbands at the edges for opposing spin directions (red and blue colors and arrows). For a confining potential $V_\perp(r)$ in the direction perpendicular to the edges and a free electron motion parallel to them, the energy states can be described by the following Hamiltonian[25]

$$H = \frac{p_\perp^2}{2m^*} + V_\perp(r) + \frac{\hbar^2}{2m^*}k_\parallel^2. \quad (2)$$

Along the edge, electrons propagate as plane waves and states are labeled as $k_{//}$. Perpendicular to the edge, we retain the momentum operator $p_\perp$. The effective electron mass $m^*$ is in the order of 0.2 $m_e$.[14] Generally, at a small spin-orbit coupling, the 1D energy levels are doubly spin degenerate. Therefore, the conductance is expected to be $2 \cdot e^2/h$ within the first subband independent of the Fermi level. However, at the surface of $Bi_2Se_3$ and $BiSbTe_3$, the parabolic bands are modified by a Rashba spin-orbit interaction because of the structural asymmetry in the materials.[13,14] The spin-orbit interaction can be expressed as[25]

$$H_{\text{Rashba}} = \alpha(\sigma_x k_\parallel - \sigma_y p_\perp/\hbar), \quad (3)$$

with a spin-orbit coupling strength α ranging from 0.3 eVÅ to 1.3 eVÅ.[13,14,28] The effect of spin-orbit coupling on the energy bands can be understood intuitively by the following argument. The first term $\alpha \cdot \sigma_x k_\parallel$ effectively shifts states that have opposite spin projections by $k_{SO} = \alpha m^*/\hbar$ with respect to each other. In this case, there exists a rigid spin-momentum locking with the in-plane spin perpendicular to the momentum along the edge direction. However, large values of $k_{SO}$ lead to crossing points of the first and second one-dimensional subbands.[25] At these crossings, the second term $\alpha/\hbar \cdot \sigma_y p_\perp$ results in a mixing of the states,[25,26] which might be further supported by an asymmetric confining potential.[34] Due to the mixing, an anti-crossing occurs in the energy dispersion. In this case, a quasi-helical regime was suggested within a certain energy range.[25,26] When the Fermi level is within this energy range (green line in Fig. 3b), there exist two counter-propagating states with an almost opposite spin direction at a large $k$-vector (blue and red arrows). At a small $k$-vector, the subbands are mixed, and the spin polarization vanishes. As a consequence, these states are not protected from backscattering,[26] and in a simple picture, the resulting conductance is expected to be $1 \cdot e^2/h$. Experimentally, we observe both $G_{photo} = 1 \cdot e^2/h$ and $G_{photo} = 2 \cdot e^2/h$ under different conditions. In Fig. 3c, we depict $G_{photo}$ of the $Bi_2Se_3$-film, which is initially capped with a protective 30 nm thick selenium layer (sketch in Fig. 3d). The circuit geometry is identical to the one depicted in Fig. 1a. With this protective selenium cap, we observe a conductance quantization centered at 1.97 $e^2/h$ (FWHM = 0.47 $e^2/h$). Tuning the Fermi level, $G_{photo}$ stays constant. When we heat the sample under atmospheric conditions, we can remove the protective selenium layer (Fig. 3e). After this procedure, we find that the conductance is quantized at 1.06 $e^2/h$ (FWHM = 0.09 $e^2/h$) for $V_{gate}$ = +30 V voltage and the value increases towards ~2 · $e^2/h$ for $V_{gate}$ = +110V (Fig. 3f). This step-wise increase is consistent with the energy level diagram in Fig. 3b. In addition, we interpret the findings with the selenium cap on top, such that initially, a weakly confined surface inversion layer is formed, which exhibits only a small spin splitting (Fig. 3a). The weak confinement results in a rather broad distribution with FWHM = 0.47 $e^2/h$, which is larger than the calculated noise limit of 0.21 $e^2/h$ for this specific measurement. Without the protective cap, however, the surface inversion layer's confinement and Rashba splitting increase.[13,14,28]

Figures 3g and 3h depict the measured conductance (circles) as a function of gate voltage together with the numerically computed values (solid lines) as a function of the Fermi level. The numerical results are shown for varying values of the Thomas-Fermi screening length $l_{TF}$ ranging from 1 nm (dark grey) to 7 nm (light grey). For a weak spin-orbit coupling, the calculated conductance only shows a single plateau at $2 \cdot e^2/h$, and it consistently describes the data for the $Bi_2Se_3$-film capped with Se. For a strong spin-orbit coupling, the conductance additionally exhibits an intermediate plateau at $1 \cdot e^2/h$. The simulations and the experimental data are in very good qualitative agreement when assuming a screening length $2\,\text{nm} < l_{TF} < 7\,\text{nm}$.[30] In particular, a dip in the conductance evolves for large $l_{TF}$ and $\alpha$ (Supplementary Figure 4), which we find consistently in the experimental data. In the simulations, the dip stems from off-diagonal elements of the complex scattering matrix. A further peculiar observation is that in most cases, the edges show a certain direction of photocurrent and therefore a specific conductance sign (Supplementary Note 3). We observe that the sign can depend on $V_{gate}$ (Supplementary Figure 5) and therefore, on the energy of the propagating electron. Hereby, the sign is given by the specific potential landscape along the quantum wire. This landscape is dominated by the presence of the defect states at the edges. The presence and importance of optically occupied defect states is experimentally verified by the sublinear power dependence in Fig. 2c (Supplementary Figure 3b).

Our numerical calculations suggest that the conductance $1 \cdot e^2/h$ results from an amplitude mixing of all states at the Fermi-energy at the presence of defect states. This mixing scenario is different to the existence of pure helical states.[25,26] Instead, a simultaneous scattering of states at large and small $k$-vectors seems to reduce the transmission of the spin-polarized states at large *k*-vectors by about a factor of one half. We note that future theory work needs to include electron-electron interactions and lateral spin-orbit field-effects, which are supposed to give rise to a spin-polarized transport in quasi one-dimensional geometries.[35,36] The experimental data, however, clearly suggest a robust quantum transport at $1 \cdot e^2/h$. The observation of the quantized edge transport is independent of the geometry of the circuits and material (Supplementary Figure 1). We note that the edge transport is not detected directly. It is locally excited by the laser and macroscopically detected at the contacts. Therefore, bulk states might be involved in the macroscopic current flow as well. In particular, we observe that the sub-micron photocurrent patterns – such as depicted with very small amplitude in Fig. 1b – are still observable in the center of the films also at positive $V_{gate}$ (Fig. 1c). At this voltage, the films are metallic. However, the observed edge transport can be excited even non-locally; i.e. at positions where no current flow occurs between the contacts (Supplementary Figure 1). This non-locality corroborates the existence of an edge channel. In our understanding, the laser gives rise to a local voltage signal which is macroscopically detected (Fig. 1d). Without a laser excitation, the edge states need to be contacted independently of the bulk states to enable a sufficiently large transport signal, e.g. by sophisticated metal contacts positioned at the edges. Finally, we note that the edge photocurrent persists up to ≈25 K in a temperature series (Supplementary Figures 6 and 7). The quantized value of $G_{photo}$ is independent of *T*, as anticipated for a macroscopic quantum transport.

## Methods

**Scanning photocurrent microscopy**

Photocurrent experiments are perfomed using a confocal laser scanning microscope with a diffraction limited spatial resolution FWHM ≈ 1 μm. We use pulsed excitation with ≈100 ps pulses at a 40 MHz repetition rate and a photon energy 1.5 eV. The optical excitation energy is larger than the band gap of $Bi_2Se_3$ (≈ 0.3 eV), but much smaller than the band gap of the $SrTiO_3$ substrate. The results are reproduced with cw-excitation. The photocurrent $I_{photo}$ at each excitation position is measured via a low-impedance transconductance amplifier at the drain contact. The transconductance amplifier provides a virtual ground potential at drain. At the same time, a voltage source maintains zero bias ($V_{SD}$ = 0 V) between source and drain contact. The photovoltage is read out via a high impedance differential amplifier (1 TΩ input impedance), which draws negligible input current. A global backgate $V_{gate}$ adjusts the charge carrier density. All other contacts except source and drain are floating. We use DC photocurrent/-voltage measurments, since kHz modulated lock-in measurements suffer from spurious background signals. The experiments are conducted at temperatures $T$ = 4.2 K – 33 K in a He atmosphere at 10 mbar pressure.

**Growth and lithography parameters of $Bi_2Se_3$ and $BiSbTe_3$ films**

The $Bi_2Se_3$ ($BiSbTe_3$) thin films are grown in a home-made molecular beam epitaxy (MBE) system with a base pressure better than $1 \times 10^{-10}$ mbar. Prior to the growth, the $SrTiO_3$(111) substrates are boiled in deionized water at 85 °C for 50 minutes, then annealed in pure $O_2$ environment at 1050 °C for 2 hours in order to obtain smooth surfaces. During the growth, high purity Bi (99.997%) and Se (99.999%) for $Bi_2Se_3$ and Bi (99.999%), Sb (99.999%) and Te (99.999%) for $BiSbTe_3$ are evaporated from Knudsen cells. A quartz crystal thickness monitor is used to calibrate the flux rate. A Bi/Se (Bi/Te) flux ratio of about 1:10 is kept to ensure Se-rich (Te-rich) growth condition. The Bi:Sb composition ratio is varied by adjusting Sb flux. The deposition rate is about 0.25 nm/min. The substrate temperature is maintained at 360°C throughout film growth. The $Bi_2Se_3$ samples are in-situ capped by 30 nm Se. After being taken out of the MBE chamber, the films are patterned into 50 μm wide Hall bars by standard photolithography, followed by reactive ion etching with an Ar flow rate of 40 sccm and a coil power of 80 W. The etching rate is about 1 nm/s. Cr/Au (3 nm/30 nm) layers are then deposited on the back of the $SrTiO_3$ substrates and the top surfaces of the films to serve as back-gate electrodes and ohmic contacts to the $Bi_2Se_3$ ($BiSbTe_3$) thin films, respectively. The Se-capping layer is removed by heating the samples to 120°C for 20 min.

**Acknowledgements**

We thank M. Wimmer for supporting us in the spin-selective numerical simulations, and V. Khrapai, A. Efros, J.P. Kotthaus, and U. Zülicke for very valuable discussions. This work was supported by the DFG via SPP 1666 (grant HO 3324/8), ERC Grant NanoREAL (n°306754), the "Center of NanoScience (CeNS)" in Munich, and the Munich Quantum Center (MQC).

**Author Contribution**

C.K. and A.W.H. designed the experiments. X.H., K.W., Y.L. fabricated the films. C.K. and P.S. performed the experiments. All authors analysed the data and wrote the manuscript.

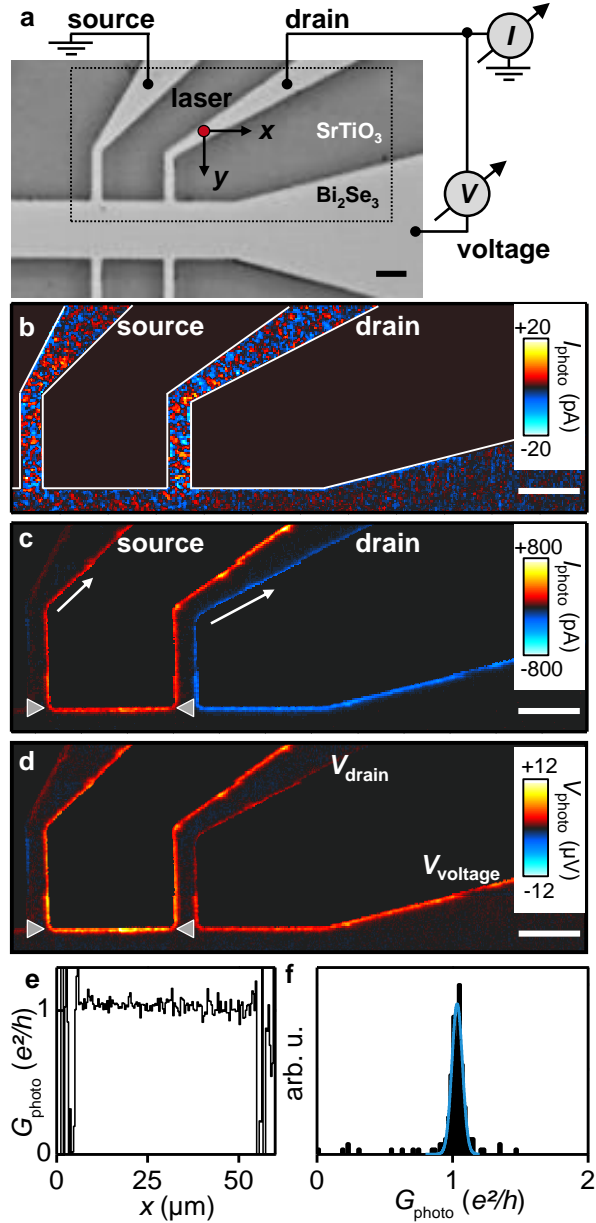

**Fig. 1. Macroscopic edge channels in Bi$_2$Se$_3$ and conductance quantization.** (a) Microscope image of a Hall bar circuit fabricated from 10 nm thick Bi$_2$Se$_3$ on SrTiO$_3$. A focused laser beam is scanned across the circuit, and the resulting photocurrent at each excitation position is measured between unbiased source and drain contacts. A differential amplifier senses the photovoltage between drain and a third adjacent contact. A global backgate $V_{gate}$ adjusts the charge carrier density. (b) At a negative $V_{gate} = -40$ V, the photocurrent is dominated by potential fluctuations. The depicted area corresponds to the dotted rectangle in Fig. 1a. (c) and (d), Simultaneously recorded photocurrent and photovoltage maps. At $V_{gate} = +30$ V, a dominant photoresponse occurs at the lateral circuit edges (dashed lines). Positive (negative) current $I_{photo}$ corresponds to an electron flow into source (drain), as indicated by arrows. (e) Photoconductance $G_{photo}$ along the edge marked by triangles in (b) and (c). (f) Normalized histogram of $G_{photo}$ peaks at 1.03 $e^2/h$ indicating one-dimensional quantum transport. Scale bars, 25 μm.

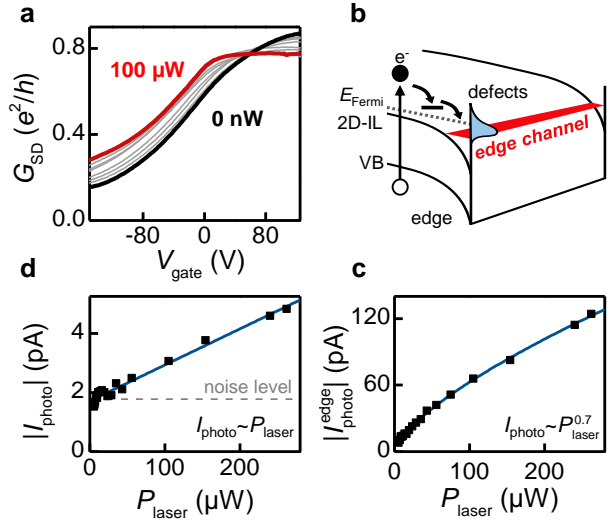

**Fig. 2. Edge photocurrents.** (a) Two-terminal source-drain conductance $G_{SD}$ without laser illumination (black line) and with laser illumination at increasing power from 0 nW up to 100 µW (red line). At $V_{gate} \geq +20$ V the source-drain conductance develops a plateau under illumination. (b) At the top surface of $Bi_2Se_3$, a two-dimensional inversion layer is formed (2D-IL). Photogenerated electrons relax via intermediate states into defects localized at the circuit edge (light blue), which leads to a lateral band bending due to a local pinning of Fermi level $E_{Fermi}$. Hereby, one-dimensional edge channels form along the lateral boundary of the circuits (red). (c) The sub-linear dependence of the edge photocurrent amplitude on laser intensity is characteristic for involved trap states. (d) In the fluctuation dominated regime, the average photocurrent amplitude $|I_{photo}|$ is linear in laser intensity $P_{laser}$.

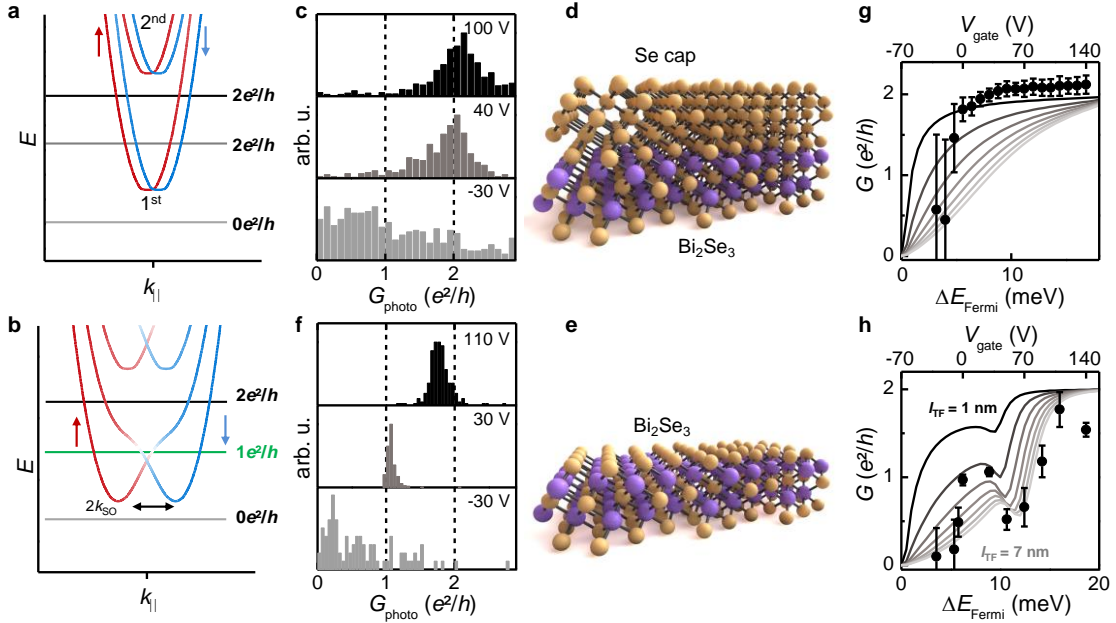

**Fig. 3. Conductance quantization at $e^2/h$.** (a) One-dimensional subbands at a small Rashba-splitting. First and second subbands are spin degenerate. The arrows and the colors blue and red indicate opposite spin projections. The conductance is quantized at $2 \cdot e^2/h$. (b) Schematics of one-dimensional subbands with strong Rashba-splitting. Near the crossing points of first and second subbands, an inter-subband mixing leads to partial spin polarization at small $k_\parallel$. At large $k_\parallel$, the subbands are almost fully spin-polarized (red and blue arrows). The solid lines indicate different Fermi levels and the corresponding conductance quantizations. (c) Histograms of $G_{photo}$ for a $Bi_2Se_3$ film with a protective selenium layer. The photoconductance is quantized at $2 \cdot e^2/h$ independent of the gate voltage. (d) and (e) Sketch of the $Bi_2Se_3$ film with and without a Se capping layer. (f) Histograms of $G_{photo}$ for varying gate voltage after removal of the protective Se layer. For $V_{gate} = 30$ V, $G_{photo}$ is quantized at $1 \cdot e^2/h$. At $V_{gate} = 110$ V, $G_{photo}$ approaches $2 \cdot e^2/h$. (g) and (h) Measured conductance $G(V_{gate})$ for capped (uncapped) $Bi_2Se_3$ film (circles) and numerically calculated conductance $G$ at the presence of impurity scattering (solid lines) for weak (strong) spin-orbit coupling. The calculations are shown for different values of the screening length $1$ nm $< l_{TF} < 7$ nm.

# Quantized one-dimensional edge channels with strong spin-orbit coupling in 3D topological insulators

- Supplementary information -

Christoph Kastl,[1,2] Paul Seifert,[1,2] Xiaoyue He,[3] Kehui Wu,[3] Yongqing Li,[3] Alexander Holleitner[1,2*]

[1] Walter Schottky Institut and Physics department, Technical University of Munich, Am Coulombwall 4a, 85748 Garching, Germany.

[2] Nanosystems Initiative Munich (NIM), Schellingstr. 4, 80799 München, Germany.

[3] Institute of Physics, Chinese Academy of Sciences, Beijing 100190, China.

[*]holleitner@wsi.tum.de

**Supplementary Figure 1**

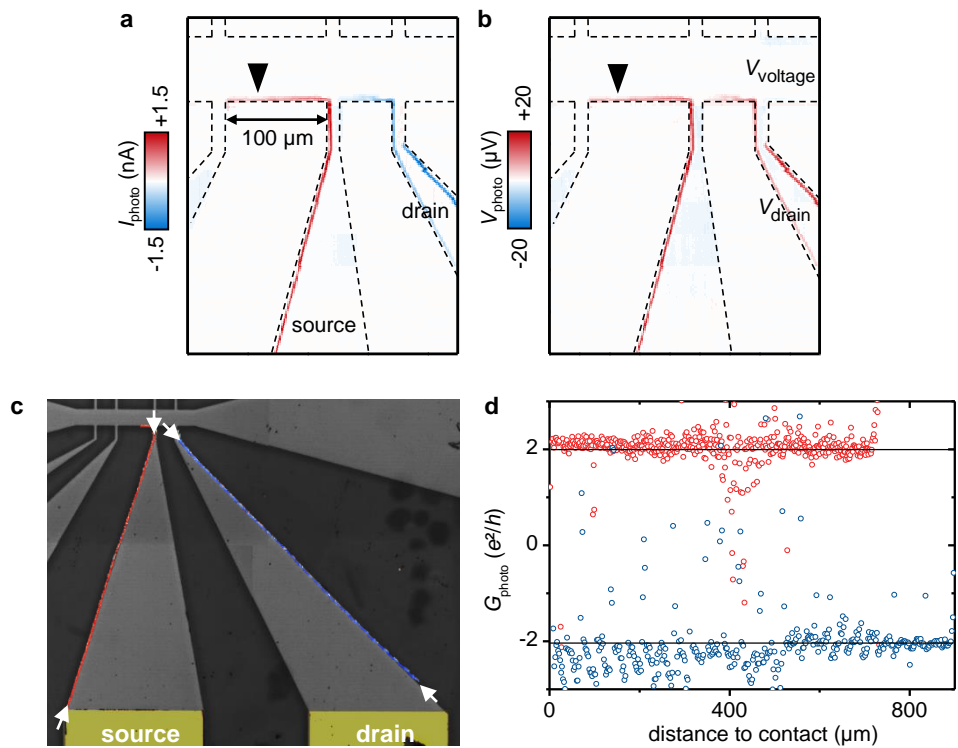

**Supplementary Figure 1. Non-local, macroscopic transport in edge channels of $Bi_2Se_3$ films.** Simultaneously recorded photocurrent **(a)** and photovoltage **(b)** maps of a 10 nm thin $Bi_2Se_3$ film. The current is measured between the contacts labelled source and drain. The voltage is measured between the contacts labelled $V_{voltage}$ and $V_{drain}$. Peculiarly, the edge transport can be excited non-locally, i.e. at positions where no current flow occurs between the source and drain contacts [triangles in (a) and (b)]. **(c)** Large scale optical microscope image showing the macroscopic contact geometry to the source and drain metal contacts. The calculated photoconductance at two edges is overlaid onto the image. The arrows highlight the two edges. **(d)** Quantized photoconductance along the two edge channels in (c) for a quantization of $2 \cdot e^2/h$. The quantization is independent of the width of the contacts, i.e. the geometry of the circuit. All measurements at $V_{gate} = +100$ V and 4.2 K.

## Supplementary Figure 2

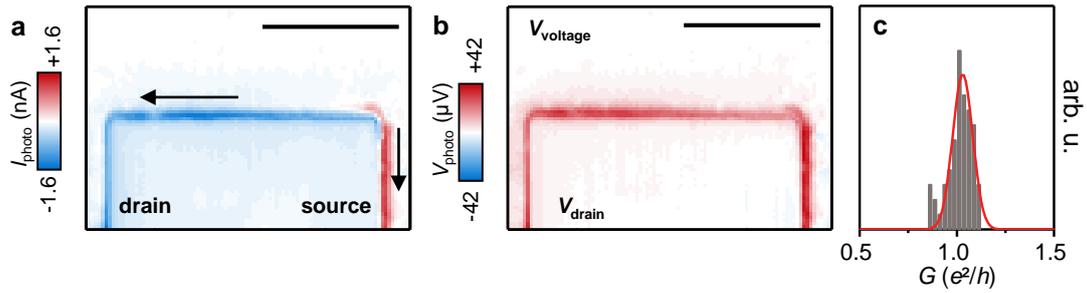

**Supplementary Figure 2. Macroscopic one-dimensional, quantized edge states in $(Bi_{0.5}Sb_{0.5})_2Te_3$ films.** Simultaneously recorded photocurrent (**a**) and photovoltage (**b**) maps of a 15 nm thin $(Bi_{0.5}Sb_{0.5})_2Te_3$ film ($V_{gate} = +40$ V). Positive (negative) current $I_{photo}$ corresponds to an electron flow into source (drain) as indicated by the arrows. The circuitry is equivalent to Fig. 1 in the main manuscript. Scale bars, 25 μm. (**c**) The histogram of the conductance $G = I_{photo}/V_{photo}$ shows a quantization at 1.03 $e^2/h$.

## Supplementary Figure 3

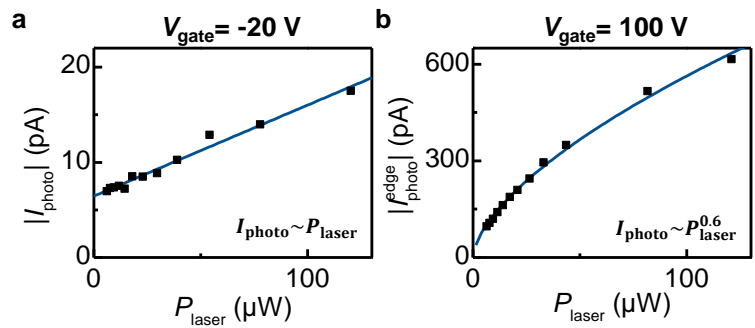

**Supplementary Figure 3. Dependence of photocurrent on laser intensity for $(Bi_{0.5}Sb_{0.5})_2Te_3$.** (**a**) Photocurrent amplitude $|I_{photo}|$ as function of laser intensity $P_{laser}$ for 15 nm thin $(Bi_{0.5}Sb_{0.5})_2Te_3$ film. The gate voltage ($V_{gate} = -20$ V) adjusts the Fermi level to the potential fluctuation dominated regime. $|I_{photo}|$ depends linearly on laser intensity. Solid line is a linear fit to the data. The vertical offset is the finite noise level. (**b**) Edge photocurrent amplitude $|I_{photo}^{edge}|$ as function of laser intensity $P_{laser}$. The gate voltage ($V_{gate} = +100$ V) adjusts the Fermi level to the edge current regime. $|I_{photo}^{edge}|$ depends sublinearly on laser intensity. Solid line is a power law fit $|I_{photo}^{edge}| = I_0 \cdot P_{laser}^p$ with $p = 0.6$.

## Supplementary Figure 4

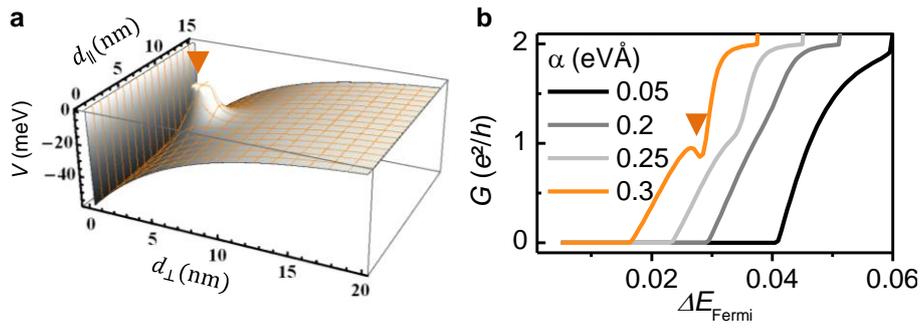

**Supplementary Figure 4 Numerical simulations of the edge channel conductance at the presence of Coulomb scattering.** (a) Assumed asymmetric potential well carrying the 1D channels with a Coulomb potential (indicated by triangle) placed at the centre of the potential well. (b) Numerically computed conductance $G(\Delta E_{Fermi})$ for the assumed potential in (d) at varying $\alpha$ in the range of $0.05\ \text{eVÅ}$ and $0.3\ \text{eVÅ}$. For an increasing $\alpha$, an additional conductance step develops (triangle).

## Supplementary Figure 5

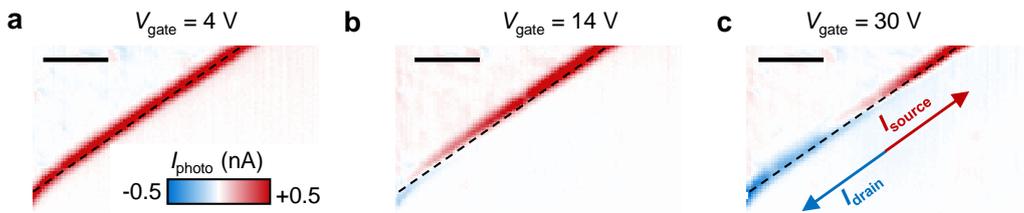

**Supplementary Figure 5. Sign of edge current along the edge of a $(Bi_{0.5}Sb_{0.5})_2Te_3$ film.** Photocurrent for $V_{gate} = +4$ V (a), $V_{gate} = +14$ V, $V_{gate} = +30$ V. Scale bars, 5 μm.

## Supplementary Figure 6

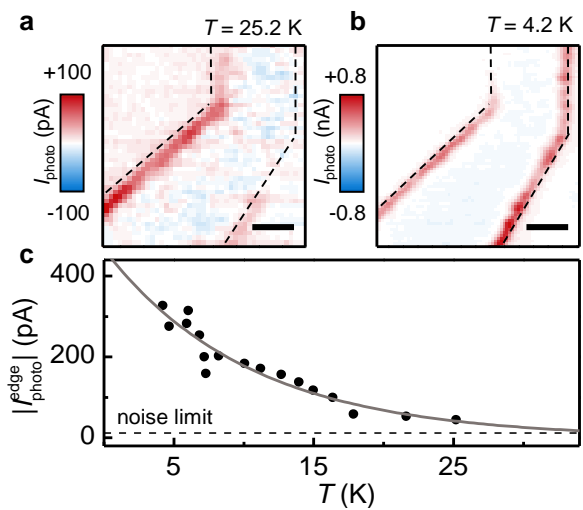

**Supplementary Figure 6. $Bi_2Se_3$: Temperature dependence of the edge photocurrent.** (a) Photocurrent map at $T = 25.2$ K. The edge current persists up to 25.2 K, but with a reduced amplitude. (b) Photocurrent map at $T = 4.2$ K. Scale bar, 5 μm. (c) Averaged amplitude of edge photocurrent $\left|I_{photo}^{edge}\right|$ as function of temperature $T$. The solid line is guide to the eye.

**Supplementary Figure 7**

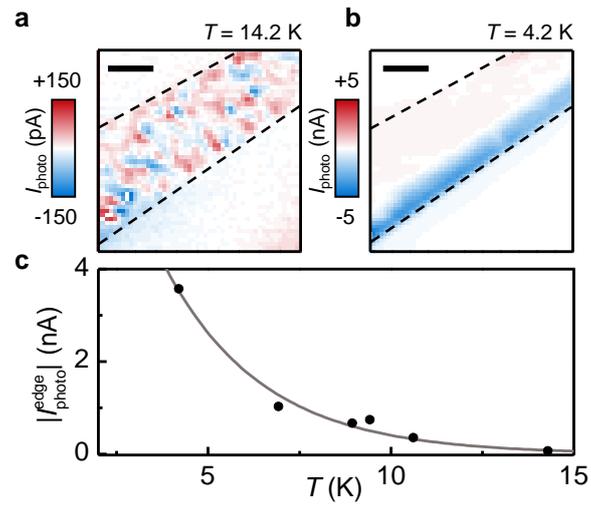

**Supplementary Figure 7 ($Bi_{0.5}Sb_{0.5})_2Te_3$: Temperature dependence of edge photocurrent.** (**a**) Photocurrent map at $T$ = 14.2 K. The edge current vanishes at 14.2 K. (**b**) Photocurrent map at $T$ = 4.2 K. Scale bar, 5 µm. (**c**) Averaged amplitude of edge photocurrent $\left|I_{photo}^{edge}\right|$ as function of temperature $T$. The solid line is a guide to the eye.

## Supplementary Note 1

**Experimentally determined width of the conductance quantization:** In the presentation of Figs. 1e and 1f of the main manuscript, we show that the edge states have a well-defined conductance with a mean value $G_{\text{photo}} = (1.03 \pm 0.005)e^2/h$, and a sharp distribution with a full width half maximum $\Delta G_{\text{photo}} = 0.091\ e^2/h$. Here, we show that the broadening is only limited by the experimental uncertainty of the current and voltage measurement. To this end, we determine the experimental noise of the current $\Delta I_{\text{photo}}$ and voltage $\Delta V_{\text{photo}}$ by calculating the standard deviation of 100 data points at positions where the photocurrent and photovoltage signals vanish, i.e. for excitation outside of the Hall bar circuit. We find $\Delta I_{\text{photo}} = 4.47$ pA and $\Delta V_{\text{photo}} = 0.20$ µV. Next, we determine the mean values $I_{\text{photo}} = 235.7$ pA and $V_{\text{photo}} = 5.889$ µV for the data in figure 1. From this, we can calculate the broadening of the conductance $\Delta G_{\text{photo}}^{\text{calc}}$ due to noise to be

$$\Delta G_{\text{photo}}^{\text{calc}} = 2.35 \sqrt{\left(\frac{1}{V_{\text{photo}}}\Delta I_{\text{photo}}\right)^2 + \left(\frac{I_{\text{photo}}}{V_{\text{photo}}^2}\Delta V_{\text{photo}}\right)^2} = 0.096 e^2/h$$

Since $\Delta G_{\text{photo}}^{\text{calc}}$ agrees very well with $\Delta G_{\text{photo}}$, we conclude that the intrinsic broadening, e.g. due to temperature, is negligible.

## Supplementary Note 2

**Numerical simulations:** We perform numerical simulations of the quantum transport through a one-dimensional edge channel with spin-orbit coupling using the Kwant code.[1] We model the edge channel by an asymmetric potential well (Supplementary Fig. 4a). Parallel to the edge, the system is translationally invariant. Perpendicular to the edge, we introduce a quantum confinement via a downwards band bending of 60 meV. Following ref. 3, such a potential is expected to be screened on a length scale of approximately 7 nm in $Bi_2Se_3$. This approach and the values are motivated by a recent experiment of us, where we demonstrate that $Bi_2Se_3$- and $BiSbTe_3$- based circuits exhibit a lateral band bending on the order of tens of meV at the edges.[2] For the simulation, we discretize the continuous Hamiltonian

$$H = -\frac{\hbar^2}{2m^*}\left(\partial_x^2 + \partial_y^2\right) + V(x,y),$$

including the spin orbit coupling term

$$H_{\text{Rashba}} = -i\alpha\left(\partial_x \sigma_y - \partial_y \sigma_x\right)$$

on a square lattice of (75×100) points at a lattice spacing of 2 Å, which corresponds to a region of 15 nm × 20 nm (Supplementary Fig. 4a). For the effective mass, we assume $m^* = 0.2 \cdot m_e$ following ref [4]. The main figures 3a and 3b show the one-dimensional subbands calculated for this potential well at a spin-orbit coupling strength of $\alpha = 0.05$ eVÅ and $\alpha = 0.3$ eVÅ, respectively. For small values of $\alpha$, the effect of spin-orbit coupling is to shift the subbands of opposite spin projections by $k_{\text{SO}} = \frac{\alpha m^*}{\hbar^2}$ (main Figure 3a). If $\alpha$ is large enough such that the energy correction due to $H_{\text{Rashba}}$ is comparable to the subband spacing, the energy bands are strongly modified (main Figure 3b). As stated in the main manuscript, the resulting band structure can be understood intuitively as an avoided crossing and therefore mixing of the first and second one-dimensional subbands. As a consequence, the spin polarization vanishes near these virtual crossing points.[5,6]

To account for the effect of scattering on the quantum transport through this edge channel, we place a screened ionized impurity at the center of the channel (triangle in Supplementary Fig. 4a). We assume $\kappa = 50$ for the effective background dielectric constant of the $Bi_2Se_3$ following ref. 7. We then calculate the energy dependent transmission of the quantum well states through the scattering region in terms of a simulated conductance $G_{sim}$. The results are displayed in the Supplementary Fig. 4b for different values of the spin-orbit coupling strength $\alpha$. At $\alpha = 0.05$ eVÅ, the conductance shows a steplike increase from 0 to $2 \cdot e^2/h$, which is smeared out by scattering off the impurity. With an increasing value of $\alpha$, a plateau-like feature develops at approximately $1 \cdot e^2/h$ (downwards triangle in the Supplementary Fig. 4b).

## Supplementary Note 3

**Sign of photocurrent and photovoltage:** In the following section, we discuss the relation between the sign of the photocurrent and the polarity of the photovoltage. On the one hand, we observe both positive and negative photocurrent signs (Supplementary Fig. 2a). A positive (negative) sign of $I_{photo}$ corresponds to an electron flow into the source (drain) contact, as indicated by the solid arrows. On the other hand, the concurrently measured photovoltage is always positive (Supplementary Fig. 2b). The latter can be understood as follows. The photovoltage ($V_{drain} - V_{voltage}$) is measured between the contacts labelled $V_{drain}$ and $V_{voltage}$. First, we consider the electron current into drain (negative current in Supplementary Fig. 2a). At the metallic contact, the electron flow equilibrates and creates a chemical potential drop $\Delta\mu < 0$ at drain. Hence, a positive voltage ($V_{voltage} - V_{drain}$) > 0 is measured. Next, we consider the electron current into source (positive current in Supplementary Fig. 2a). Again, at the metallic contact, the electron flow equilibrates and creates a chemical potential drop $\Delta\mu < 0$ at source. The voltage probe draws no current and the source-drain voltage is set to $V_{SD} = 0$ V ($V_{source} = V_{drain}$) by the external voltage source. In turn, we measure a positive voltage ($V_{voltage} - V_{drain}$) > 0.

A peculiar observation is that in most cases, the edge states show a certain direction of photocurrent and therefore conductance sign. Apparently, one factor which determines the direction of $I_{photo}$ is the geometry of the circuit. For instance, in Supplementary Fig. 2a, the photocurrent changes sign at a kink of the edge, and the current flow near the kink points towards the nearer contact. However, we find that the sign of $I_{photo}$ can also depend on $V_{gate}$ (Supplementary Figure 5) and therefore, on the energy of the propagating electrons. In other words, the sign is given by the effective potential landscape along the quantum wire which is dominated by the presence of the defect states at the edges.